# New Phase Transition of Lanthanum at High Pressure

Tianxiao Liang,[1] Zihan Zhang,[1] Hao Song,[1] and Defang Duan[1]*

[1] College of Physics, Jilin University, Changchun 130012, China

Lanthanum (La), the first member of the rare-earth elements, recently aroused strong interest due to its unique superhydride with superconducting properties. Although there is much theoretical and experimental work about phase transitions and superconductivity in metallic La, we got a new body-centred tetragonal ($bct$) phase in metallic La with space group $I4/mmm$ at 190 GPa exchanging from face-centered cubic ($fcc$) phase in previous work, which expanded the phase transition sequence. The $bct$ phase shows an abnormal packing way that turned to non-closed packing at high pressure. And more detailed properties of the new phase are discussed.

## Introduction

Lanthanum (La) is the first element of the lanthanide series, and the unique physical and chemical properties of lanthanide series have attracted interest of researchers for many years. Under high pressure the regular trivalent lanthanide elements and Sc, Y show a well-known close packed series of structures: $(d)hcp \rightarrow$ Sm-type$(fcc) \rightarrow dfcc \rightarrow fcc$ [1], and a $bcc$ structure of La is favorable at about 1,100 ~ 1,200 K, which is near the melting point [2-4]. It's also known that the structure stability in rare-earth metals is controlled by $d$ −electron number which increases under pressure due to $s - d$ transfer [1].

Previous experiments indicate that at ambient conditions lanthanum takes either the stable $\alpha$-La double-$hcp$ ($dhcp$) or the metastable $\beta$-La ($fcc$) structure, and phase transitions occurred at 2.5 GPa for $dhcp$ to $fcc$, then to distorted-$fcc(dfcc)$ at about 5.4 GPa, and finally return to $fcc$ at about 60 GPa [5-7]. Further calculation by Nixon *et al.* shows that the $dhcp$ transforms to the $fcc$ near 2.2 GPa and shifts to distorted$-fcc$ ($dfcc$) at about 2.5 GPa, then returns to the $fcc$ phase near 90 GPa [8], which shows a little different experimental results. In 2018, Geballe *et al* provided more details about pressure-volume relationship for $fcc$ phase at the range of 50 ~ 180 GPa [9]. Recently, Chen *et al* declared that they discovered a new $dfcc$-La with space group $Fmmm$ from

78 GPa to 140 GPa, which has tiny errors of the lattice parameters with $fcc$-La, and superconductivity survives in La with critical temperature $T_c$ of 2.2 K at 140 GPa [10] in experimental work by piston-cylinder diamond anvil cell (DAC), under non-hydrostatic pressure conditions. In Chen's work, lattice parameters of La-$Fmmm$ are very similar to lattice parameters of $fcc$ phase under the corresponding pressures, with almost the same volumes. And La-$Fmmm$ may be synthesized by unbalanced pressures from different axis and the distortion of La-$fcc$ may not occur at highly hydrostatic pressure. In our study, at properly hydrostatic pressure, relative to the previous works, we did more on theoretical calculation of La, and we finally made the phase series expand to body-centred tetragonal ($bct$) with a new phase La-$I4/mmm$ under higher pressure. The new phase shows an abnomal packing method, which is not closest packing method.

## Computational Details

Structure searches of La were performed with *ab initio* calculations as implemented in the AIRSS (Ab Initio Random Structure Searching) codes [11, 12] and CALYPSO structure prediction method [13, 14]. The CASTEP code [15] was used for the AIRSS searches. The VASP (Vienna ab initio simulation packages) code [16] was used to optimize crystal structures and calculate the electronic properties, where the Perdew–Burke–Ernzerhof (PBE) [17] of generalized gradient approximation (GGA) [18] was chosen for the exchange-correlation functional. And the all-electron projector-augmented wave method (PAW) [19] was performed. Kinetic cutoff energy of 800 eV was selected. and Monkhorst-Pack **k** meshes with grid spacing of $2\pi \times 0.03 Å^{-1}$ were then adopted to ensure the enthalpy converges to less than 1 meV/atom. And we used the all-electron full-potential linearized augmented plane wave (FP-LAPW) method within the framework of density functional theory as implemented in WIEN2k code [20]. The electronion interaction was described by projector-augmented-wave potentials with the $5s^2 5p^6 5d^1 6s^2$ configurations treated as valence electrons for La. The phonon calculations were carried out by the PHONOPY code [21], which showed a good agreement to those computed using the Quantum-ESPRESSO code [22]. Lattice dynamics and electron–phonon coupling (EPC) were characterized by density functional perturbation theory as implemented in the Quantum-ESPRESSO code, and superconducting properties were calculated using density functional perturbation theory (DFPT) [23], performed with the Quantum-ESPRESSO code as well. The first-order potential perturbation and dynamical matrices were calculated using DFPT on an irreducible $4 \times 4 \times 5$ $\Gamma$-centered **q**-point mesh. Ultra-soft pseudopotential [24] were used with a kinetic energy cutoff of 80 Ry, and the charge density is integrated on a $15 \times 15 \times 17$ $\Gamma$-centered **k**-point mesh. A Methfessel-Paxton first-order smearing of 0.02 Ry was applied. The Bader charge analysis [25] was used to determine charge transfer, and the electron localization function (ELF) [26] was used to describe and visualize chemical bonds in molecules and solids.

## Results and Discussions

Full structure predictions were firstly performed within 20,000 structures using random structure searches based on both two codes at the range of 50 to 250 GPa, with the pressure step of 50 GPa. We have identified a new phase of La with space group $I4/mmm$, which is different from La-$Fm\overline{3}m$ [5-7, 9] or La-$Fmmm$ [10] in the previous work. La atoms occupied $2a$ sites in $I4/mmm$ lattice and $4a$ sites in $Fm\overline{3}m$.

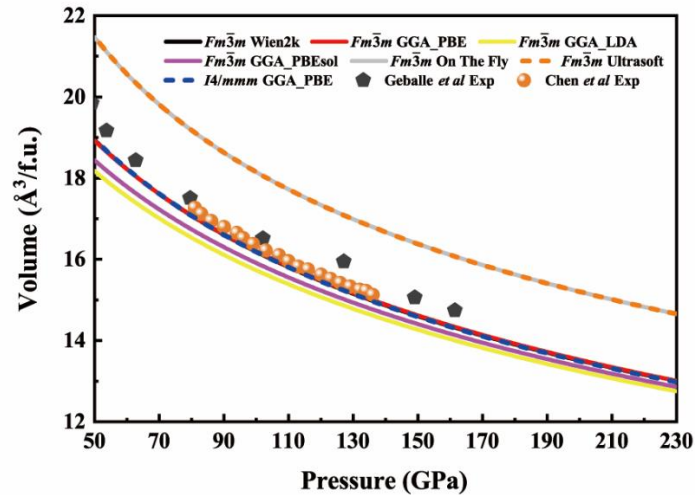

FIG 1. Volumes as a function of pressures for La-$Fm\bar{3}m$ in red solid line and La-$I4/mmm$ in blue dash line are calculated by using PAW potential in VASP calculations. Other methods are also preformed relative to full-potential in WIEN2K calculations in black solid line. Calculated results are relative to Geballe *et al* s' experimental work [10] in brown pentagons, and Chen *et al* s' experimental work [10] in filled orange balls.

Volumes as a function of pressures (P-V) for La-$Fm\bar{3}m$ calculated by using PAW potential in VASP calculations, where the GGA_PBE functional was performed (red solid line in Fig. 1), and full-potential in WIEN2K calculations (black solid line in Fig. 1). And P-V curve of La-$I4/mmm$ was given identical results in Fig. 1 as well (blue dash line). Other methods with different potentials and exchange-correlation functionals are also performed. In order to determine the parameters of the equation of states (EOS), the obtained P-V data were fitted by the third-order Birch-Murnaghan equation [27]:

$$P = \frac{3B_0}{2}\left[\left(\frac{V_0}{V}\right)^{\frac{7}{3}} - \left(\frac{V_0}{V}\right)^{\frac{5}{3}}\right]\left\{1 + \frac{3}{4}(B_0' - 4)\left[\left(\frac{V_0}{V}\right)^{\frac{2}{3}} - 1\right]\right\} \quad (1)$$

where $V_0$ is the equilibrium cell volume, $B_0$ is the bulk modulus, and $B_0'$ is the derivative of bulk modulus with respect to pressure. The detailed EOS curves relative to Exp. results are plotted in Fig. 1. The fitted parameters for La-$Fm\bar{3}m$ are $V_0 = 26.15 Å^3$, $B_0 = 77.73$ GPa, and $B_0' = 4.09$. And the fitted parameters for La-$I4/mmm$ are $V_0 = 27.70 Å^3$, $B_0 = 58.06$ GPa, and $B_0' = 4.27$. There is almost no volumetric changing for La-$Fm\bar{3}m$ and La-$I4/mmm$. And we can see that our volumetric changing calculations fit the experimental results properly at the pressure range of 100 ~ 140 GPa. Our P-V results are also well fitting with experimental results in Fig. 1. At higher pressure above 190 GPa, there is no experimental datas. The P-V results suggest the PAW method performing with GGA_PBE functional is reliable in the pressure range of current work. From our calculations, the tendencies of volumetric changing for La-$Fm\bar{3}m$ and La-$I4/mmm$ are consistent.

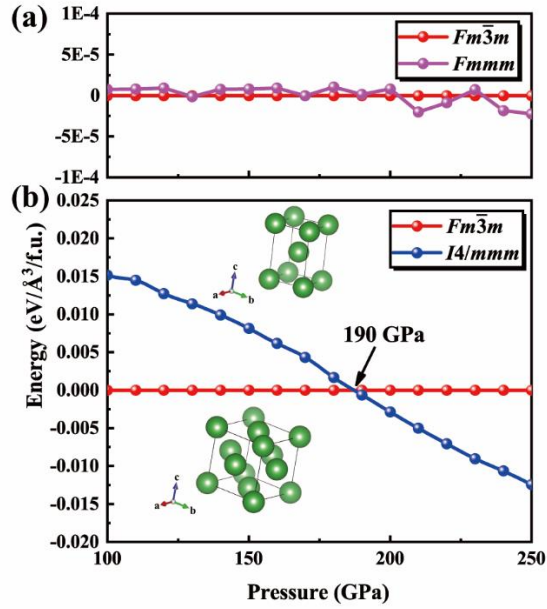

FIG 2. Calculated enthalpies as a function of pressure for La-$I4/mmm$ relative to the $Fm\bar{3}m$ phase calculated by using full-potential WIEN2K package. The phase-transition pressures was calculated to become ~190 GPa.

Thermodynamic calculation in Fig. 2(a) indicates that $Fmmm$ and $Fm\bar{3}m$ have almost the same enthalpy, and they are thermodynamically competitive phases to each other. Geometry optimizations are performed with full structural relaxation including atomic positions and lattice constants to detect whether structural phase transformation is induced by pressure. We believe that La-$Fmmm$ phase is synthesize from La-$Fm\bar{3}m$ by non-hydrostatic pressure, and the real phase transition should occur at higher pressure, and there should be a new phase. Calculation in Fig. 2(b) indicates that La-$Fm\bar{3}m$ transforms to the La-$I4/mmm$ at about 190 GPa.

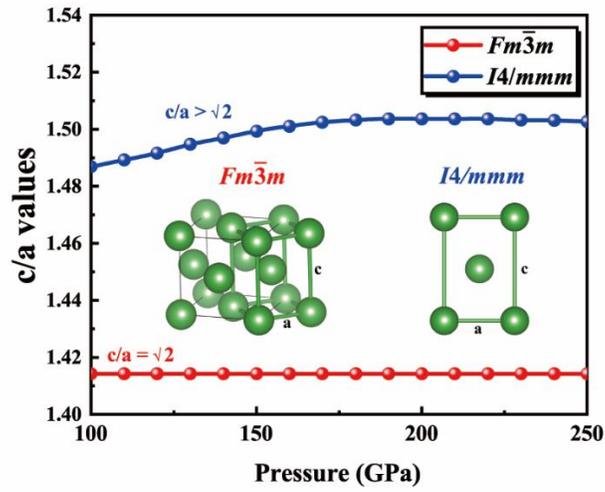

FIG 3. c/a values for $I4/mmm$ phase relative to the $Fm\bar{3}m$ phase.

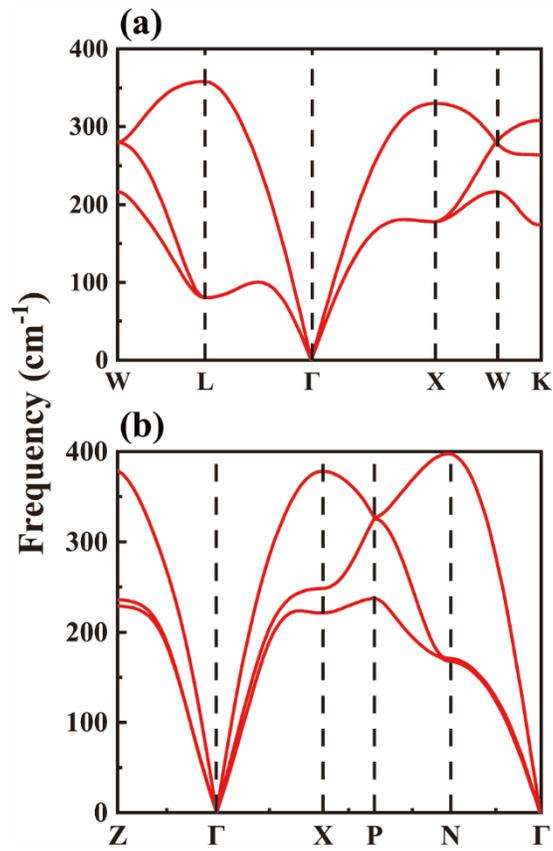

FIG 4. The calculated phonon dispersion relations for La-$Fm\bar{3}m$ at (a) 100 GPa, (b) 150 GPa, and (c) for La-$I4/mmm$ at 200 GPa.

It is clearly seen that the lattice parameters changed continuously with pressures in Fig. 3. At 200 GPa, lattice parameters of La-$I4/mmm$ are $a = b = 2.623 Å$ and $c = 3.939 Å$, where $\alpha = \beta = \gamma = 90°$. If we redefine the crystal lattice of La-$Fm\overline{3}m$ in body-centred tetragonal packing method like La-$I4/mmm$ listed in Fig. 3, it must have $c/a = \sqrt{2}$ for La-$Fm\overline{3}m$. And we got $c/a > \sqrt{2}$ for La-$I4/mmm$ in Fig. 3 with blue ball symbols and solid line. This result significantly indicated that La-$I4/mmm$ and La-$Fm\overline{3}m$ were structurally two different phases. Coordination numbers of La-$I4/mmm$ and La-$Fm\overline{3}m$ are 8 and 12. The phase transition leads to be an abnormally loose packing method under high pressure, rather than close packing method like hexagonal close packed or face-centered cubic packing. It is an interesting phenomenon.

With dynamic calculations shown in Fig. 4, the absence of imaginary frequency in the phonon spectra indicates the structures of La-$Fm\overline{3}m$ and La-$I4/mmm$ are dynamically stable at corresponding pressure ranges. In our further investigation, the superconductive transition temperature $T_c$ was estimated through the Allen-Dynes-modified McMillan equation with correction factors [28]:

TABLE I. Calculated electronic density of states ($N_{Ef}$), logarithmic average frequency ($\omega_{\log}$), the electron-phonon coupling parameter ($\lambda$), and $T_c$ of La at Fermi level at 200 GPa at different pressures.

| Phase | P (GPa) | $\mu^*$ | $N_{Ef}$ (States/eV/Å$^3$) | $\omega_{\log}$ (K) | $\lambda$ | $T_c$ (K) |
|---|---|---|---|---|---|---|
| $Fm\overline{3}m$ | 60 | 0.10/0.13 | 6.435 | 130.06 | 1.03 | 9.53/8.24 |
| $Fm\overline{3}m$ | 100 | 0.10/0.13 | 5.515 | 167.95 | 0.70 | 5.85/4.84 |
| $Fm\overline{3}m$ | 150 | 0.10/0.13 | 5.293 | 227.18 | 0.33 | 1.88/1.20 |
| $I4/mmm$ | 200 | 0.10/0.13 | 5.146 | 437.93 | 0.20 | 0.00 |

$$T_c = \frac{\omega_{\log}}{1.2} \exp\left[\frac{-1.04(1+\lambda)}{\lambda-\mu^*-0.62\lambda\mu^*}\right] \quad (2)$$

where $\lambda$ is the EPC constant and the $\omega_{\log}$ prefactor is a properly defined logarithmic average frequency suggested by Allen and Dynes, while $\mu^*$ is Coulomb pseudopotential, which is a parameter accounting for the effective Coulomb repulsion. The EPC parameter $\lambda$ and logarithmic average frequency $\omega_{\log}$ are [28] :

$$\lambda = \int_0^{\omega_{\max}} \frac{2\alpha^2 F(\omega)}{\omega} d\omega \qquad (3)$$

$$\omega_{\log} = \exp\left(\frac{2}{\lambda} \int_0^{\omega_{\max}} \frac{2\alpha^2 F(\omega)\ln\omega}{\omega} d\omega\right) \qquad (4)$$

As shown in Table 1, the superconducting transition temperature $T_c$ of La-$Fm\overline{3}m$ is 1.9 K at 150 GPa at $\mu^* = 0.1$ using Perdew-Burke-Ernzerhof (PBE) pseudopotentials and obtaining $\lambda = 0.33$ and $\omega_{\log} = 227.18$ K, which is similar to 2.2 K at 140 GPa by Chen *et al*. At 100 GPa, we got $T_c = 5.85$ K, obtaining $\lambda = 0.70$ and $\omega_{\log} = 167.95$ K. Both Tissen *et al* [7] and Chen *et al* got experimental $T_c$ of 10.5 K at 50 GPa and close to the calculated value (9.3 K) [10]. And we got $T_c = 9.53$ K at 60 GPa, obtaining $\lambda = 1.03$ and $\omega_{\log} = 130.06$ K at $\mu^* = 0.1$. We also simulated the influence of non-hydrostatic pressure on lattice distortion, which made the La lattice distort slightly. But the small errors of three lattice parameters $a$, $b$, and $c$ of La-$Fmmm$ compared with La-$Fm\overline{3}m$ had almost little effect on the superconducting results. It indicated that La-$Fmmm$ got by Chen *et al* was mostly like La-$Fm\overline{3}m$ lattice slightly distorted by non-hydrostatic pressure, rather than a new phase. And new phase changed from should be La-$I4/mmm$, which was gotten at 190 GPa in our work. Unfortunately, the $T_c$ of La-$I4/mmm$ tended to be 0 K, and it seems to be no superconductivity at higher pressure.

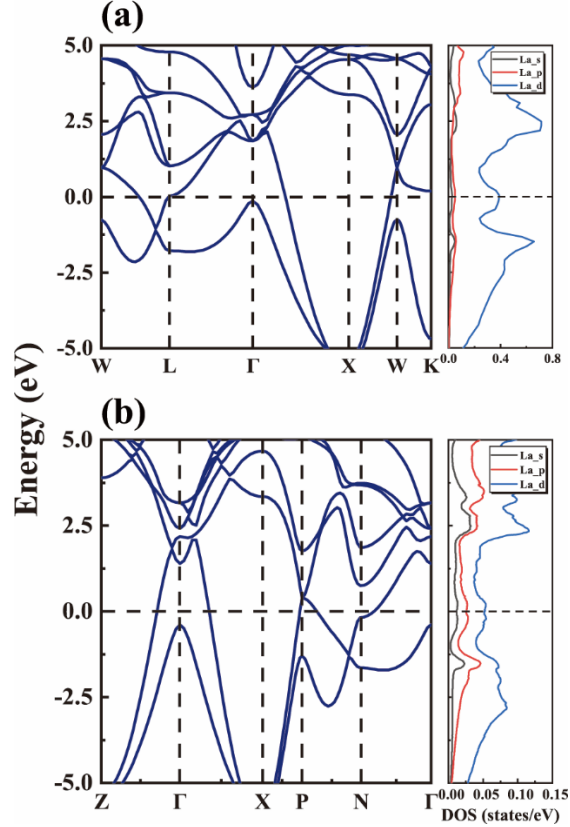

FIG 5. The calculated electronic band structure and projected density of states for La-$Fm\bar{3}m$ at (a) 100 GPa, (b) 150 GPa, and (c) for La-$I4/mmm$ at 200 GPa.

Fig. 5 are the calculated band structure (left panel) and projected density of states (right panel) for La-$Fm\bar{3}m$ and La-$I4/mmm$. As we can see, La-$Fm\bar{3}m$ and La-$I4/mmm$ are conductors at high pressure, and $d$ electrons donate the most at Fermi surfaces for both of the two phases. It is obvious that the contribution of $d$ electron significantly reduces after the phase transition occurred. And the Bader analysis naturally showed the metallicity of La-$I4/mmm$ phase as well. We believe that the decreasing of $d$ electrons' contribution in La lattice prevents further compact packing and leads to this abnormally packing methods, which is different from Sc or Y in IIIB group at high pressure [1].

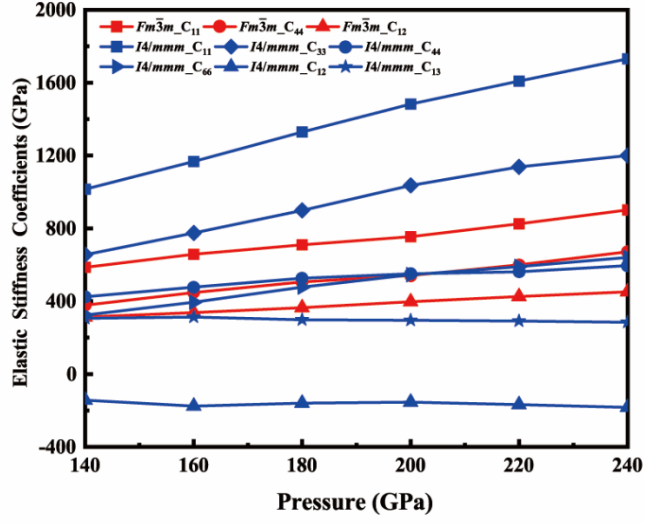

FIG 6. Pressure dependence of elastic constants in Born stability criteria for La-$Fm\bar{3}m$ in red and La-$I4/mmm$ in blue.

To explore more about La at high pressure, the elastic constants are calculated in Fig. 6 to describe its response to an applied stress or the stress required to maintain a given deformation. The cubic crystal system has the simplest form of elastic matrix, with only 3 independent constants: $C_{11}$, $C_{12}$, and $C_{44}$ (Laue class is $m\bar{3}m$) [29]:

$$\mathbb{C}_{\text{cubic}} = \begin{bmatrix} C_{11} & C_{12} & C_{12} & & & \\ C_{12} & C_{11} & C_{12} & & & \\ C_{12} & C_{12} & C_{44} & & & \\ & & & C_{44} & & \\ & & & & C_{44} & \\ & & & & & C_{44} \end{bmatrix} \quad (5)$$

The widely used Born stability criteria for cubic crystal system are [30]:

$$C_{11} > |C_{12}|;\ C_{11} + 2C_{12} > 0;\ C_{44} > 0 \quad (6)$$

Crystals of the tetragonal class thus have 6 independent elastic constants, which are $C_{11}$, $C_{12}$, $C_{13}$, $C_{33}$, $C_{44}$, and $C_{66}$ to form the elastic matrix (Laue class is $4/mmm$) [29]:

$$\mathbb{C}_{\text{tetra}} = \begin{bmatrix} C_{11} & C_{12} & C_{13} & & & \\ C_{12} & C_{11} & C_{13} & & & \\ C_{13} & C_{13} & C_{33} & & & \\ & & & C_{44} & & \\ & & & & C_{44} & \\ & & & & & C_{66} \end{bmatrix} \quad (7)$$

And Born stability criteria for tetragonal crystal system are [30]:

$$\begin{cases} C_{11} > |C_{12}|; 2C_{13}^2 < C_{33}(C_{11}+C_{12}); \\ C_{44} > 0; C_{66} > 0 \end{cases} \quad (8)$$

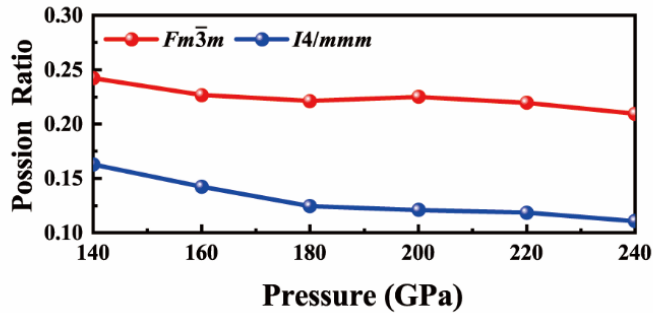

FIG 7. Calculated Poisson's ratios of La-$Fm\bar{3}m$ in red and La-$I4/mmm$ in blue.

As shown in Fig. 6, $C_{11}$, $C_{12}$, and $C_{44}$ of La-$Fm\bar{3}m$ increase linearly and slowly, and show no sign of softening. $C_{11}$ and $C_{33}$ of La-$I4/mmm$ increase rapidly, and $C_{66}$ show a significant growth with the increasing pressure as well. But $C_{12}$, $C_{13}$, and $C_{44}$ of La-$I4/mmm$ are almost unchanged with the increasing pressure. Datas of La-$Fm\bar{3}m$ and La-$I4/mmm$ in Fig. 6 satisfy the Born stability criteria from Equ. 6 and Equ. 8, it indicates that our results show that La-$Fm\bar{3}m$ and La-$I4/mmm$ satisfy the Born stability criteria in the pressure range from 140 to 220 GPa. Calculated Poisson's ratios and modulus of La-$Fm\bar{3}m$ and La-$I4/mmm$ are shown in Fig. 7 and Fig. 8. The Poisson's ratio of a stable, isotropic, linear elastic material must be between -1.0 and +0.5 because of the requirement for the bulk modulus, shear modulus and Young's modulus to have positive values, and structures with lower Poisson's ratios are more compressible [31]. As shown in Fig. 7, Poisson's ratio curve of La-$Fm\bar{3}m$ has a very gentle fluctuation at the pressure range of 180 ~ 200 GPa. And phase transition occurred at 190 GPa, which is reflected by Poisson's ratio curve transformation. See Fig. 8, La-$I4/mmm$ has similar but larger bulk modulus compared with La-$Fm\overline{3}m$'s bulk modulus. And La-$I4/mmm$ has significantly larger shear modulus and Young's modulus than those of La-$Fm\overline{3}m$. It indicates that La-$I4/mmm$ has a significantly deformation compared with La-$Fm\overline{3}m$.

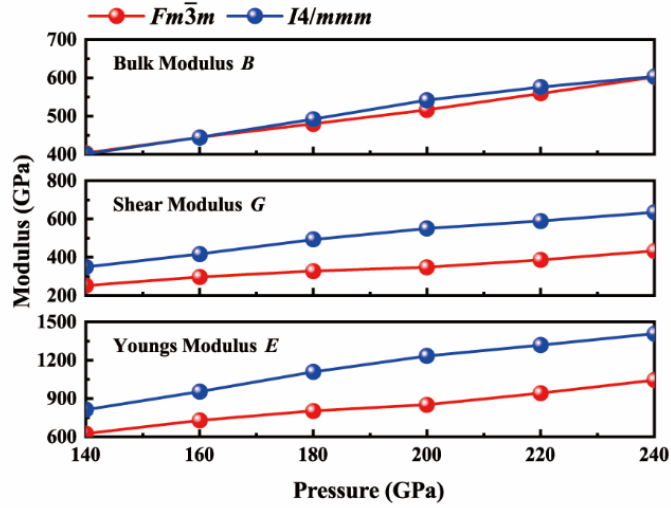

FIG 8. Calculated modulus of La-$Fm\overline{3}m$ in red and La-$I4/mmm$ in blue.

## Conclusions

To conclude, we have predicted the existence of new structure of La: La-$I4/mmm$ at 190 GPa transformed from La-$Fm\overline{3}m$. The abnormal packing method of lanthanum tends to be a loose method rather than ordinary closest packing methods. Our calculations for phononic, electronic, and mechanical properties have revealed this abnormal pressure-induced phase transition. We showed that La-$I4/mmm$ has no superconductivity at 200 GPa. The calculated elastic stiffness coefficients $C_{ij}$ of La structures satisfy the Born stability criteria and it indicates that La-$Fm\overline{3}m$ and La-$I4/mmm$ are elastically stable at high pressure. By further investigation of Poisson's ratios and modulus, we believe that the pressure leads to a significant phase transition from La-$Fm\overline{3}m$ to La-$I4/mmm$. As for the reason of the abnormal packing method, we suppose that it should be considered as the $s-d$ transfer of La. The contribution of $d$ electron significantly reduces as the pressure increases. From the above discussion there should be a new phase for La at high pressure. And this result may be promoted in other lanthanides or actinides.

## Acknowledgement

This work was supported by National Natural Science Foundation of China (Nos. 11674122 and 11704143). Parts of calculations were performed in the High Performance Computing Center (HPCC) of Jilin University and TianHe- 1 (A) at the National Supercomputer Center in Tianjin.

## References

1. Altmann, S.L., C.A. Coulson, and W. Hume-Rothery, *On the relation between bond hybrids and the metallic structures.* Proceedings of the Royal Society of London. Series A. Mathematical and Physical Sciences, 1957. **240**(1221): p. 145-159.
2. Finnemore, D.K., et al., *Superconductivity in pure La and La-Gd.* Physical Review, 1965. **137**(2A): p. A550.
3. Homan, C., R.K. McCrone, and E. Whalley, *High Pressure in Science and Technology.* Parts I, II, III, 1984.
4. Jarlborg, T., et al., *Resistivity, bandstructure and superconductivity of DHCP and FCC La under pressure.* Journal of Physics: Condensed Matter, 1989. **1**(44): p. 8407.
5. Porsch, F. and W.B. Holzapfel, *Novel reentrant high pressure phase transtion in lanthanum.* Physical review letters, 1993. **70**(26): p. 4087.
6. Tissen, V.G., et al., *Effect of pressure on the superconducting T c of lanthanum.* Physical Review B, 1996. **53** (13): p. 8238.
7. Wittig, J., *Note on the superconductivity of uranium at high pressure.* Zeitschrift f{\"u}r Physik B Condensed Matter, 1975. **22** (2): p. 139-142.
8. Nixon, L.W., D.A. Papaconstantopoulos, and M.J. Mehl, *Electronic structure and superconducting properties of lanthanum.* Physical Review B, 2008. **78** (21): p. 214510.
9. Geballe, Z.M., et al., *Synthesis and stability of lanthanum superhydrides.* Angewandte Chemie, 2018. **130** (3): p. 696-700.
10. Chen, W., et al., *Superconductivity and equation of state of lanthanum at megabar pressures.* 2020.

11. Pickard, C.J. and R.J. Needs, *High-pressure phases of silane.* Physical Review Letters, 2006. **97** (4). 045504.
12. Pickard, C.J. and R.J. Needs, *Ab initio random structure searching.* Journal of Physics: Condensed Matter, 2011. **23** (5): p. 053201.
13. Wang, Y., et al., *Crystal structure prediction via particle-swarm optimization.* Physical Review B, 2010. **82** (9): p. 094116.
14. Wang, Y., et al., *CALYPSO: A method for crystal structure prediction.* Computer Physics Communications, 2012. **183** (10): p. 2063-2070.
15. Segall, M.D., et al., *First-principles simulation: ideas, illustrations and the CASTEP code.* Journal of Physics: Condensed Matter, 2002. **14** (11): p. 2717.
16. Kresse, G. and J. Hafner, *14251); g. kresse, j. furthmuller.* Phys. Rev. B, 1996. **54** : p. 11169.
17. Perdew, J.P., K. Burke, and M. Ernzerhof, *Generalized gradient approximation made simple.* Physical review letters, 1996. **77** (18): p. 3865.
18. Perdew, J.P. and Y. Wang, *Pair-distribution function and its coupling-constant average for the spin-polarized electron gas.* Physical Review B, 1992. **46** (20): p. 12947.
19. Blöch, P.E., *Projector augmented-wave method.* Physical review B, 1994. **50** (24): p. 17953.
20. Blaha, P., et al., *Full-potential, linearized augmented plane wave programs for crystalline systems.* Computer Physics Communications, 1990. **59** (2): p. 399-415.
21. Togo, A., F. Oba, and I. Tanaka, *First-principles calculations of the ferroelastic transition between rutile-type and CaCl 2-type SiO 2 at high pressures.* Physical Review B, 2008. **78** (13): p. 134106.
22. Giannozzi, P., et al., *QUANTUM ESPRESSO: a modular and open-source software project for quantum simulations of materials.* Journal of physics: Condensed matter, 2009. **21** (39): p. 395502.
23. Baroni, S., et al., *Phonons and related crystal properties from density-functional perturbation theory.* Reviews of Modern Physics, 2001. **73** (2): p. 515.
24. Vanderbilt, D., *Soft self-consistent pseudopotentials in a generalized eigenvalue formalism.* Physical review B, 1990. **41** (11): p. 7892.
25. Bader, R.F.W., *International Series of Monographs on Chemistry.* Atoms in Molecules, A Quantum Theory, 1990. **22**.
26. Becke, A.D. and K.E. Edgecombe, *A simple measure of electron localization in atomic and molecular systems.* The Journal of chemical physics, 1990. **92** (9): p. 5397-5403.
27. Birch, F., *Finite elastic strain of cubic crystals.* Physical review, 1947. **71** (11): p. 809.
28. Allen, P.B. and R.C. Dynes, *Transition temperature of strong-coupled superconductors reanalyzed.* Physical Review B, 1975. **12** (3): p. 905.
29. Mouhat, F.e.l. and F.c.o.-X. Coudert, *Necessary and sufficient elastic stability conditions in various crystal systems.* Physical review B, 2014. **90** (22): p. 224104.
30. Born, M. and K. Huang, *Theory of Crystal Lattices, Clarendon.* 1956. p..
31. Gercek, H., *Poisson's ratio values for rocks.* International Journal of Rock Mechanics and Mining Sciences, 2007. **44** (1): p. 1-13.